\documentclass[12pt,preprint]{aastex}
\received{2003 October 17}
\accepted{2004 June 24}
\slugcomment{Accepted for publication in the ApJ: June 24, 2004}
\begin{document}
\title{Deep Near Infrared Imaging toward Vela Molecular Ridge C\\ - 1: A Remarkable Embedded Cluster in RCW 36 - } 

\shorttitle{Deep NIR Imaging toward VMR C 1}
\shortauthors{Baba et al.}

\author{D.~Baba\altaffilmark{1,2}, T.~Nagata\altaffilmark{1}, T.~Nagayama\altaffilmark{1}, C.~Nagashima\altaffilmark{1}, D.~Kato\altaffilmark{1}, M.~Kurita\altaffilmark{1}, S.~Sato\altaffilmark{1}, Y.~Nakajima\altaffilmark{3}, M.~Tamura\altaffilmark{3}, H.~Nakaya\altaffilmark{4}, and  K.~Sugitani\altaffilmark{5}}

\altaffiltext{1}{Department of Physics, Nagoya University, Chikusa-ku, Nagoya 464-8602, Japan}

\altaffiltext{2}{e-mail: baba@z.phys.nagoya-u.ac.jp}

\altaffiltext{3}{National Astronomical Observatory of Japan, Mitaka, Tokyo 181-8588, Japan}

\altaffiltext{4}{Subaru Telescope, National Astronomical Observatory of Japan, Hilo, HI 96720}

\altaffiltext{5}{Institute of Natural Sciences, Nagoya City University, Mizuho-ku, Nagoya 467-8501, Japan}

\begin{abstract}
We present deep near-infrared ($J, H, K_{S}$) images toward an embedded cluster which lies in a C$^{18}$O clump in the cloud C of the Vela Molecular Ridge.
This cluster has at least $\sim$ 350 members and a radius of $\sim$ 0.5 pc. 
The stellar surface number density is approximately 3000 stars pc$^{-2}$ in the central 0.1 pc $\times$ 0.1 pc region of the cluster. 
This is much higher than most of the young clusters within 1 kpc of the Sun. 
From the comparison of the luminosity function and near-infrared excess fraction with those of other embedded clusters, we estimate that the age of this cluster is approximately 2-3 Myr. This cluster exhibits an excess of brighter stars in its central region, from which we conclude that the more massive stars are located near the cluster center.
\end{abstract}


\section{INTRODUCTION}
The progress of NIR array detectors provided opportunities to observe stellar contents in molecular clouds. Many studies of these regions revealed the presence of stellar clusters which are deeply embedded in molecular clouds \citep{Lada91, Carpenter93, Hodapp94}.
Statistical studies revealed that the most of young stellar objects in giant molecular clouds (GMCs) are formed in embedded clusters \citep{Lada03}.
These embedded clusters are typically found to have their most massive stars in the central region (the position whose stellar densities are the highest) \citep{Zinnecker93, Hillenbrand98}.
Several embedded clusters also shows mass segregation; more massive stars tend to be located in the central region [e.g., Trapezium \citep{Hillenbrand98}; NGC 2024, NGC 2071 \citep{Lada91}; NGC 3603 \citep{Nurnberger02}]. However, \citet{Carpenter97} pointed out that there is no evidence of mass segregation in the Mon R2 cluster. Unfortunately, the mass segregation has been investigated only for a few embedded clusters. We need more samples for statistical studies.

In this paper, we study an embedded cluster which lies in a C$^{18}$O clump identified by \citet{Yamaguchi99} in the cloud C of the Vela Molecular Ridge (VMR).
The VMR is a GMC complex, being composed of at least four GMCs (named A, B, C, and D) with individual masses exceeding 10$^{5}$ M$_{\odot}$ \citep{Murphy91}.
The clouds A, C, and D are located at 700 $\pm$ 200 pc, and the cloud B is located at $\sim$ 2 kpc \citep{Liseu92}.
The cloud C is the richest in molecular gas and the least evolved of the complex \citep{Yamaguchi99}. This embedded cluster was recently discovered by \citet{Massi03} in their NIR imaging survey toward Class I sources in VMR \citep{Liseu92, Lorenzetti93}.
Massi, et al. (2003) found that high mass stars which excite the H {\small II} region RCW 36 lie in this cluster (hereafter RCW 36 cluster).
However, due to the relatively small image dimensions, their observations were not able to cover the whole cluster.
Therefore, the fundamental properties of the RCW 36 cluster, such as the number of the stars, radius and the stellar distribution, have been still unknown. 
Our observation first provides the detailed census of the RCW 36 cluster.

\section{OBSERVATIONS AND DATA REDUCTION}
The NIR images of RCW 36 were obtained on 2002 March 9 with the NIR camera SIRIUS (Simultaneous three-color InfraRed Imager for Unbiased Survey) mounted on the IRSF (InfraRed Survey Facility) 1.4 m telescope of Nagoya University at Sutherland, South African Astronomical Observatory.
SIRIUS is equipped with three 1024 $\times$ 1024 pixel HgCdTe arrays.
Dichroic mirrors enable simultaneous observations in the $J,$ $H,$ and $K_{S}$ bands \citep{Nagashima99, Nagayama02}.
The image scale of the array is 0\arcsec .45 pixel$^{-1}$, giving a field of view of 7\arcmin .7 $\times$ 7\arcmin .7.
Ten dithered frames were observed as a set of exposures.
We observed three sets of the object and sky frames alternatively with a 30 second exposure.
We also observed one set of the object and sky frames with a 5 second exposure for bright sources.
Seeing condition was $\sim$ 1\arcsec .1 (FWHM) in the $K_{S}$ band.
We observed the standard star 9136 in the faint near-infrared standard star catalog of \citet{Persson98} for photometric calibration.

We applied the standard procedures of near-infrared array image reduction, including dark current subtraction, sky subtraction, and flat-fielding, using the IRAF (Imaging Reduction and Analysis Facility)
\footnote{IRAF is distributed by the National Optical Astronomy Observatories, which are operated by the Association of Universities for Research in Astronomy, Inc., under cooperative agreement with the National Science.} 
software package. 
Identification and photometry of point sources were performed by using the DAOPHOT packages in IRAF. 
%%We used a radius of $\sim$ 3 pixels (1\arcsec .35) in the PSF fitting in the DAOPHOT task so that the nearby stars did not affect the photometry.
In order to avoid the affection for the photometry from the nearby stars, we used a radius of $\sim$ 3 pixels (1\arcsec .35) in the PSF fitting in the DAOPHOT task.

The limiting magnitudes (at 10 $\sigma$) are estimated to be $\sim$ 19.0, 18.0, and 16.6 in the $J,$ $H,$ and $K_{S}$ bands, respectively. A total number of 823 sources are detected in the $K_{S}$ band with a photometric error of $\le$ 0.1 mag, and 610 sources are identified in the three bands with a photometric error of $\le$ 0.1 mag and with a positional agreement of $\le$ 0.5\arcsec.
In order to determine the completeness limit of the photometry, we added artificial stars to the central region of the image (dotted square in Fig.~2) with 17\arcsec~spacing.
We define the magnitudes at which 90\% of stars were detected as the completeness limits.
They are 18.2, 16.9, and 16.3 at $J,$ $H,$ and $K_{S},$ respectively.

\section{RESULTS}
	\subsection{NIR image}

In Fig.~\ref{image}, we show a composite $JHK_{S}$ color image of the RCW 36 cluster. It clearly shows that the stars concentrate at the center of the image.
The brightest star located at 8$^{h}$59$^{m}$27$^{s}$.5, -43$^{\circ}$45\arcmin 27\arcsec~(J2000) is the exciting star \citep[O8 or two O9 stars; ][]{Verma94} of RCW 36 (Massi, et al. 2003). 
At optical wavelengths, the stellar cluster including the exciting star is hidden by a dark-lane across RCW 36.
The radio continuum peak position is $\sim$ 35\arcsec~west of the exciting star \citep[shown by plus in Fig.~\ref{image};][]{Walsh98}; this coincide spatially with a bright nebula. 
An H$_{2}$O maser is detected $\sim$ 30\arcsec~west of the exciting star \citep[shown by asterisk in Fig.~\ref{image};][]{Braz82}; there is no NIR counterpart.
The existence of the H$_{2}$O maser means that star formation is still ongoing in the RCW 36 cluster.
There are several bright-rimmed clouds at $\sim$ 1\arcmin~west, $\sim$ 1\arcmin~south, and $\sim$ 2\arcmin~south-east of the exciting star.
They are likely to be dense cores which were revealed by evaporation of a molecular cloud due to UV photons from the exciting star.

	\subsection{Stellar Census in the RCW 36 Cluster}
		\subsubsection{Stellar Surface Number Density}

In Fig.~\ref{contour}, we show the contour map of the stellar surface number density (SSND) derived from 823 sources detected at $K_{S}$ with a photometric error of $\le$ 0.1 mag.
The SSND is obtained by counting stars in a 30\arcsec~$\times$ 30\arcsec~area every 15\arcsec~over the $K_{S}$ image.
At the distance of 700 pc, the unit cell of 30\arcsec~$\times$ 30\arcsec~corresponds to $\sim$ 0.1 pc $\times$ 0.1 pc.
The frequency distribution of the star counts (in stars per unit cell) can be fitted by the Poisson function with a mean of 2.0 stars per unit cell, plus the wing where excess of counts due to the central clustering appear.
The SSND shows two peaks. The northern peak coincides spatially with the exciting star and shows the highest SSND. The southern one coincides spatially with one of the bright rims at $\sim$ 1\arcmin~south of the exciting star.

		\subsubsection{Radial Profile}
In Fig.~\ref{radprof}, we show the radial profile of the SSND for the RCW 36 cluster. The SSND is measured in circular annuli, centering on the location of the peak position of the SSND.
Profiles are calculated using annuli with equal radial steps (=15\arcsec). 
The surface density profiles of embedded clusters can be fitted with 1/r profiles \citep[e.g.,][]{McCaughrean94, Jiang03, Muench03}.
They also can be fitted with King profiles \citep[e.g.,][]{Horner97, Nurnberger02, Teixeira04}.
If we can neglect the tidal radius, King profiles can be written as
\[ f(r) = \frac{f_{0}}{1+(r/r_{c})^{2}} \]
where $f_{0}$ is the central surface density and $r_{c}$ is the core radius \citep{King62}. Ideally, the radial profile should become zero at r $\rightarrow \infty$. However, in the observation of the actual clusters, the radial profile does not become zero but a constant at r $\rightarrow \infty$ due to the field stars which contribute as a background offset. Therefore, in order to take the contribution from the field stars into account, we added a constant term as a fitting parameter to fitting functions.

The data of the RCW 36 cluster can be well fitted with King profiles better than with 1/r profiles; a departure from King profiles arising around r $\sim$ 0.2 pc is due to the southern SSND peak.
King profile fitting yields the central density of $f_{0}$ = 3200 stars pc$^{-2}$ and the core radius of $r_{c}$ = 0.08 pc.
Since the column density of isothermal gas sphere in hydrostatic equilibrium has a radial dependence of r$^{-1}$ \citep{Yun91}, the 1/r dependence of the SSND radial profile in the embedded clusters is likely to be a footprint of the density profile of their parental cloud core.
On the other hand, King models describe solutions for stellar systems in dynamical equilibrium, such as globular clusters. 
The secondary peak of the SSND, which causes the departure around r $\sim$ 0.2 pc in Fig.~\ref{radprof}, might be a relic of the primordial substructure that has not fully merged into main body of the RCW 36 cluster \citep{Scally02}.
Since King profiles better fits the observed radial profile than 1/r profiles overall except for the secondary peak, the dynamical equilibrium would have been almost, if not completely, established in the RCW 36 cluster.

		\subsubsection{Determination of the Cluster Properties}
In order to obtain the fundamental properties of the RCW 36 cluster, such as the number of the stars and radius, we have to estimate the degree of the field star contamination. The frequency distribution of the star counts per unit cell (used to derive SSND in \S 3.2.1) can be fitted by the Poisson function with a mean of 2 stars per unit cell. This means that uniform background stars are distributed in the whole observed field. Hereafter, we take this value (2.0 stars/unit cell $\sim$ 190 stars pc$^{-2}$) as the background. 
This estimation is reasonable because the background value yielded by King profile fitting in the previous section is in good agreement with it.

We can estimate the number of the RCW 36 cluster members by subtracting background from 823 sources detected at $K_{S}$ with a photometric error of $\le$ 0.1 mag.
The total expected background population in the observed field is 474 [= 2.0 $\times$ 4 (stars arcmin$^{-2}$) $\times$ 7.7 $\times$ 7.7 (area of observed field)].
Then, the total number of the cluster members is estimated to be 349; we note that this estimation is a lower limit down to our limiting magnitude.
The radial profile merges with the extended background at r $\sim$ 0.5 pc (see Fig.~\ref{radprof}). We take this as the cluster radius.
Considering the background contamination, the SSND of the central 0.1 pc $\times$ 0.1 pc region of the cluster is approximately 3000 stars pc$^{-2}$; this is in good agreement with the central density yielded by King profile fitting in the previous section.
This is different from the value of $\sim$ 4800 stars pc$^{-2}$ derived by Massi, et al. (2003), but we confirmed in our data that this inconsistency is due to their unit cell size (20\arcsec~$\times$ 20\arcsec).
In table \ref{ssnd_comp}, we show the comparison of the central SSND with other embedded clusters within 1 kpc of the Sun. 
The central SSDN of the RCW 36 cluster is much higher than most of the young clusters but lower than the Trapezium cluster.

The theory borrowed from the traditional theory of low-mass stars that stars are formed through gravitational collapse of dense molecular cores can only account for those stars with masses less than 10 M$_{\odot}$, because radiation pressure of massive stars can halt the collapse and reverse the infall.
\citet{Bonnell98b} presented result of numerical experiments that the massive ($M \ge 10 M_{\odot}$) stars can be formed through coalescence of two or more lower mass stars in the center of rich, dense stellar clusters.
The most massive star in the RCW 36 cluster (O8-O9) is located at the cluster center; this might imply that it was formed through this mechanism.
 \citet{Hillenbrand95} suggested that there is a correlation of the cluster density with the maximum stellar mass in the cluster (i.e. more centrally condensed clusters have more massive stars). In fact, the most massive stars in R~CrA, IC 348, and Mon R2 are less massive than that of the RCW 36 cluster, and Trapezium  has a more massive star (see table \ref{ssnd_comp}).
However, the most massive star of NGC 2024 is approximately the same as that of the RCW 36 cluster. 
We note that the central SSND of NGC 2024 might have been estimated to be too low, because the observation toward NGC 2024 \citep{Lada91} was made with relatively small limiting magnitude (K $\sim$ 14 mag) and relatively larger pixel scale ($\sim$ 1\arcsec.3 pixel $^{-1}$).
If we assume a distance of 415 pc and mean extinction of A$_{V} \sim $ 10.5 mag toward NGC 2024 \citep{Haisch00}, the limiting magnitude corresponds to an extinction-corrected absolute magnitude of M$_{K_{A_{V}=0}} \sim$ 4.9 mag. For an assumed age of 1 Myr, their observation could detect down to $\sim$ 0.09 M$_{\odot}$ \citep{Baraffe98}.
In order to take low-mass members of NGC 2024 into account down to our observation limit [$\sim$0.04 M$_{\odot}$ assuming a distance of 700 pc, mean extinction of A$_{V} \sim$ 8.1 mag, and an age of $\sim$ 3 Myr (see below)], we use the slope of Trapezium IMF \citep{Muench02}.
The estimated central SSND of NGC 2024 ($\sim$ 1700 stars pc$^{-2}$) is still lower than that of RCW 36.
However, the larger pixel scale might affect the estimation of the central SSND more seriously, because the radial profile of the SSND for NGC 2024 decreases in the cluster center \citep[see Fig.~8 of][]{Lada91}.

	\subsection{color-color diagram}
In Fig.~\ref{2cd}, we show the J-H v.s. H-K color-color diagram for the 610 sources identified in three bands (see \S2).
Their colors are transformed to the CIT system with the color equations\footnote{\url{http://www.z.phys.nagoya-u.ac.jp/$\sim$sirius/about/color\_e.html}} (Y.~Nakajima, et al. in preparation).
The sources falling into a region of infrared excess (the right of the reddening band) in the color-color diagram are considered to be likely cluster members with optically thick, circumstellar disks.
The total of 77/349 (22\%) of the cluster members exhibit infrared excesses greater than their photometric errors (we note this excess fraction might be just a lower limit because several cluster members were not detected in three bands). 
We also estimate the mean extinction toward the RCW 36 cluster to be A$_{V} \sim$ 8.1 mag by de-reddening all sources in Fig.~\ref{2cd} back to this CTTS locus \citep{Meyer97} along the reddening vector.

	\subsection{K$_{S}$ Luminosity Function}
In Fig.~\ref{KLF}, we show the $K_{S}$ band luminosity function (hereafter KLF) for the RCW 36 cluster.
In order to obtain the KLF for the cluster members (cluster KLF), we must remove background contamination.
We use the sources which are included in the regions whose SSND is smaller than the background plus 1$\sigma$ (off-cluster region) to construct the off-cluster KLF. 
The off-cluster region is approximately the same as the region r $\ge$ 0.5 pc from the cluster center. Thus, most of the sources contained in the off-cluster region are supposed to be background stars (see Fig.~\ref{radprof}).
The off-cluster KLF is shown with open squares and dashed line in Fig.~\ref{KLF}a.
By subtracting the (scaled) off-cluster KLF from the KLF which includes both the cluster members and the field stars (raw KLF; shown in Fig.~\ref{KLF}a with filled circle and solid line), we can obtain the cluster KLF (Fig.~\ref{KLF}b). The cluster KLF has a turnover with a peak at $K_{S} \sim$ 14 mag, and then decreases down to the completeness limit.

\section{DISCUSSION}
	\subsection{Age of the RCW 36 cluster}
NIR imaging surveys revealed that, in embedded clusters, the fraction of the sources which possess NIR excess decreases with the cluster age \citep{Lada99, Haisch01b}. The NIR fraction derived from JHK imaging is approximately 50 \% for the young ($\le$ 1 Myr) embedded clusters [e.g., 50 $\pm$ 7\% in Trapezium \citep{Lada00}; 58 $\pm$ 7\% in NGC 2024 \citep{Haisch00}]. The fraction decreases to approximately 20 \% for more evolved ($\sim$ 2-3 Myr) embedded clusters [e.g., 21 $\pm$ 5 \% in IC 348 \citep{Haisch01a}; 16 $\pm$ 3 \% in NGC 2316 \citep{Teixeira04}]. The NIR excess fraction of the RCW 36 cluster (22 \%; see \S 3.3) implies that the age of the RCW 36 cluster is approximately 2-3 Myr.

A similar age is obtained by comparing the KLFs.
The comparison of the KLFs of RCW 36, Trapezium and IC 348 is shown in Fig.~\ref{KLFcomp}.
The KLFs for Trapezium and IC 348, by \citet{Muench03}, are corrected to absolute magnitudes.
Reddening corrections were performed for each KLF by shifting to brighter magnitude by each mean extinction [Av $\sim$ 8.1 mag in RCW 36 (see \S 3.3), Av $\sim$ 2.4 mag in Trapezium \citep{Prosser94}, Av $\sim$ 4.5 mag in IC 348 \citep{Lada95}]. Trapezium KLF and RCW 36 KLF are scaled to contain the same number of stars as IC 348. The shape of the KLF of RCW 36 is similar to that of Trapezium, but is shifted toward the fainter magnitude. On the other hand, the KLFs of IC 348 and RCW 36 show good agreement in shape and they have the same broad peak at similar absolute magnitudes.

\citet{Muench00} presented the result of numerical experiments that the KLFs for young clusters which have the same underlying IMF evolve in a systematic manner with increasing age; they evolve to fainter magnitudes.
The IC 348 IMF is nearly identical to the Trapezium IMF \citep{Muench03}, and they are similar to that derived for other embedded clusters, open clusters, and field stars \citep[e.g.,][and the references therein]{Lada03}. Hence, if we assume a universal IMF for the RCW 36 cluster, we infer that the RCW 36 cluster has approximately the same age as IC348, about 2-3 Myr, but is older than Trapezium.

	\subsection{Mass Segregation}
In order to examine the mass segregation in the RCW 36 cluster, we show the stellar distribution by splitting the 823 sources detected at $K_{S}$ with a photometric error of $\le$ 0.1 mag into three magnitude bins (K$_{s} <$ 12.0, 12.0 $\le$ K$_{s} < 14.0$, and 14.0 $\le$ K$_{S}$) in Fig.~\ref{masseg_dist}.
The sources of K$_{s} <$ 12.0 show high central condensation (Fig.~\ref{masseg_dist}a). The sources of 12.0 $\le$ K$_{S} <$ 14.0 also show central condensation but are less condensed than the sources of K$_{S} <$ 12.0 (Fig.~\ref{masseg_dist}b). The sources of 14.0 $\le$ K$_{S}$ show no central condensation (Fig.~\ref{masseg_dist}c);
 most of these sources are likely to be background stars because their distribution is uniform.
These figures clearly demonstrate an excess of the brighter stars in the cluster center.
This is more evident in Fig.~\ref{masseg_rad}, which shows the SSND radial profiles for three magnitude bins (the same as above). The central position of each radial profiles are the same as Fig.~\ref{radprof}.
King profile fitting for each magnitude bins yields the core radii of $r_{c}$ = 0.02 pc (Fig.~\ref{masseg_rad}a), 0.13 pc (Fig.~\ref{masseg_rad}b), and 0.24 pc (Fig.~\ref{masseg_rad}c, was not fitted well because the surface density decreases in the central annulus).

In young embedded clusters, because most members are considered to be in the pre-main sequence (PMS) phase, this tendency could have two possibilities. One is that the sources located in the cluster center are young, and the other is that they are massive. The NIR excess of the PMS stars is caused by optically thick circumstellar disks/envelopes. Since the circumstellar disks/envelopes become optically thinner with age, the PMS stars which possess NIR excess are considered to be younger than those with no NIR excess.
Therefore, if younger stars concentrate in the cluster center, the distribution of the sources which possess NIR excess should be similar to those of the brighter sources.
However, as shown in Fig.~\ref{masseg_rad}d, the radial profile for the sources which possess NIR excess does not show significant central condensation. Their King profile fitting yields the core radius of $r_{c}$ = 0.20 pc which is clearly larger than that of the sources of K$_{s} <$ 12.0. Therefore, the brightness of the sources in the central region of the RCW 36 cluster is not due to their youth, but due to their massiveness.

Mass segregation can be caused 1) by dynamical evolution and equipartition of energy after star formation and 2) by the process of star formation itself.
If mass segregation is caused by dynamical evolution, it occurs in approximately the relaxation time, $\tau_{relax}$, of the cluster \citep{Bonnell98a}.
The relaxation time is roughly
\[ \tau_{relax} \sim \frac{0.1N}{\ln N} \tau_{cross} \]
where N is the number of stars contained in the cluster and $\tau_{cross}$ is the crossing time \citep{Binney87}.
The crossing time, $\tau_{cross}$, is the time it takes a star to cross the cluster ($\tau_{cross} \sim 2R / v_{disp}$, where $R$ is the radius and $v_{disp}$ is the velocity dispersion of the cluster).
In the case of most open clusters, mass segregation is likely caused by dynamical evolution [e.g., Pleiades \citep{Pinfield98}]. 
However, mass segregation in rich and young embedded clusters, such as the Trapezium cluster, is not considered to be caused by dynamical evolution because they are too young to have dynamically evolved. Here, we estimate the relaxation time for several embedded clusters in table \ref{relax_comp}.
Since the Trapezium cluster has a small age and a large number of stars, the relaxation time is significantly larger than its age in spite of the small crossing time.
In other embedded clusters whose SSND is lower, dynamical evolution is generally very unlikely because the cluster sizes are larger, the ages are smaller, or both.
However, the relaxation time of the RCW 36 cluster is estimated to be $\sim 2.6 \times 10^{6}$ yr; this is comparable to the cluster age estimated above.
Thus, at least from our data, there is a possibility that the mass segregation in the RCW 36 cluster was caused by dynamical evolution, because the RCW 36 cluster has smaller radius, moderate number of stars, and moderate age.
However, this dose not rule out the possibility of primordial mass segregation in the RCW 36 cluster because the clusters which dynamical equilibrium are established lost all traces of their initial conditions.

\section{CONCLUSION}
We have presented deep $J, H, K_{S}$ observations of the RCW 36 cluster which lies in a C$^{18}$O clump in the cloud C of the Vela Molecular Ridge. The main conclusions are as follows.

1: The RCW 36 cluster has at least $\sim$ 350 members and a cluster radius of $\sim$ 0.5 pc, and the radial profile of the stellar surface number density can be well fitted with King profiles better than with 1/r profiles.
In the central 0.1 pc $\times$ 0.1 pc region of the RCW 36 cluster, the stellar surface number density is approximately 3000 stars pc$^{-2}$  This is much higher than most of the young clusters within 1 kpc of the Sun. 

2: We measured the NIR excess fraction of the RCW 36 cluster to be 22 \%, compared with those of other embedded clusters, and estimated the age of the RCW 36 cluster is approximately 2-3 Myr. The comparison of background corrected KLF of the RCW 36 cluster with Trapezium and IC 348 also supports this estimation. 

3: The RCW 36 cluster shows an excess of the brighter stars in its central region. From the comparison of the radial profiles with the sources of $K_{S} <$ 12.0 mag and the sources with NIR excess, we conclude that this tendency means that more massive stars are located near the cluster center.

\acknowledgments{It is pleasure to thank the staff of the South African Astronomical Observatory for their kind support during the observations.
We also would like to express our thanks to the anonymous referee for comments and suggestions, which greatly improved the scientific content of the paper.
This work is partly supported by Grant-in-Aid for Scientific Research of the Ministry of Education, Culture, Sports, Science, and Technology of Japan.}

\clearpage

\begin{figure}
%%\plotone{f1.gif}
\caption{$JHK_{S}$ three-color composite image of the RCW 36 cluster. ($J:blue,$ $H:green,$ $K_{S}:red$) obtained with IRSF/SIRIUS. North is up, and east is to the left.
The plus represents the peak position of radio continuum \citep{Walsh98}, and the asterisk represents the position of H$_{2}$O maser \citep{Braz82}. 
[Figure separately included as gif (to make smaller).]
\label{image}}
\end{figure}

\clearpage

\begin{figure}
\plotone{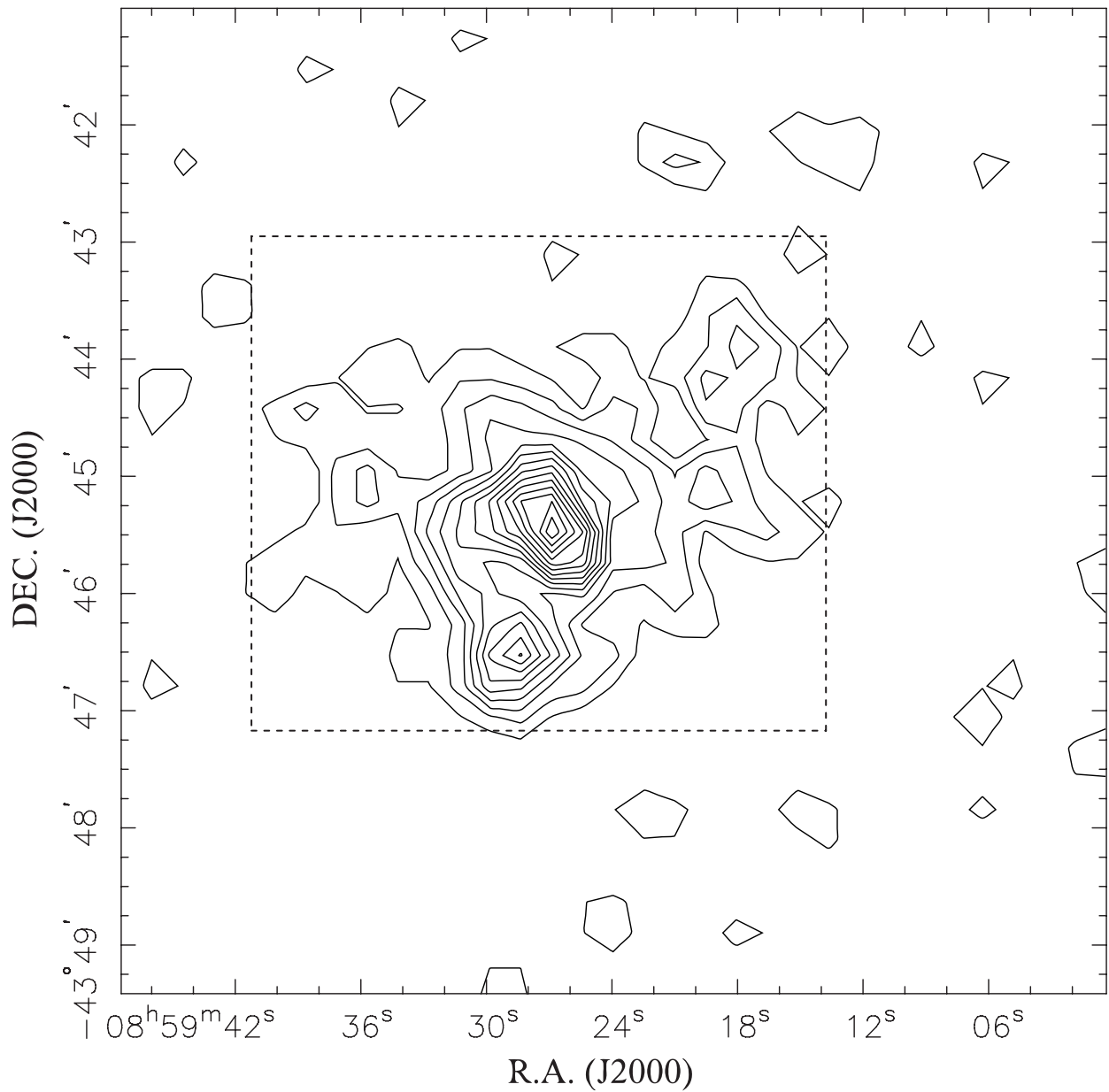}
\caption{Contour map of the stellar surface number density obtained by counting stars in a 30\arcsec~$\times$ 30\arcsec~($\sim$ 0.1 pc $\times$ 0.1 pc at 700 pc) area every 15\arcsec~over the $K_{S}$ image. The lowest contour is 400 stars pc$^{-2}$, and steps are 200 stars pc$^{-2}$. Dotted square is the region used to determine the completeness limits (see \S2).
\label{contour}}
\end{figure}

\clearpage

\begin{figure}
\plotone{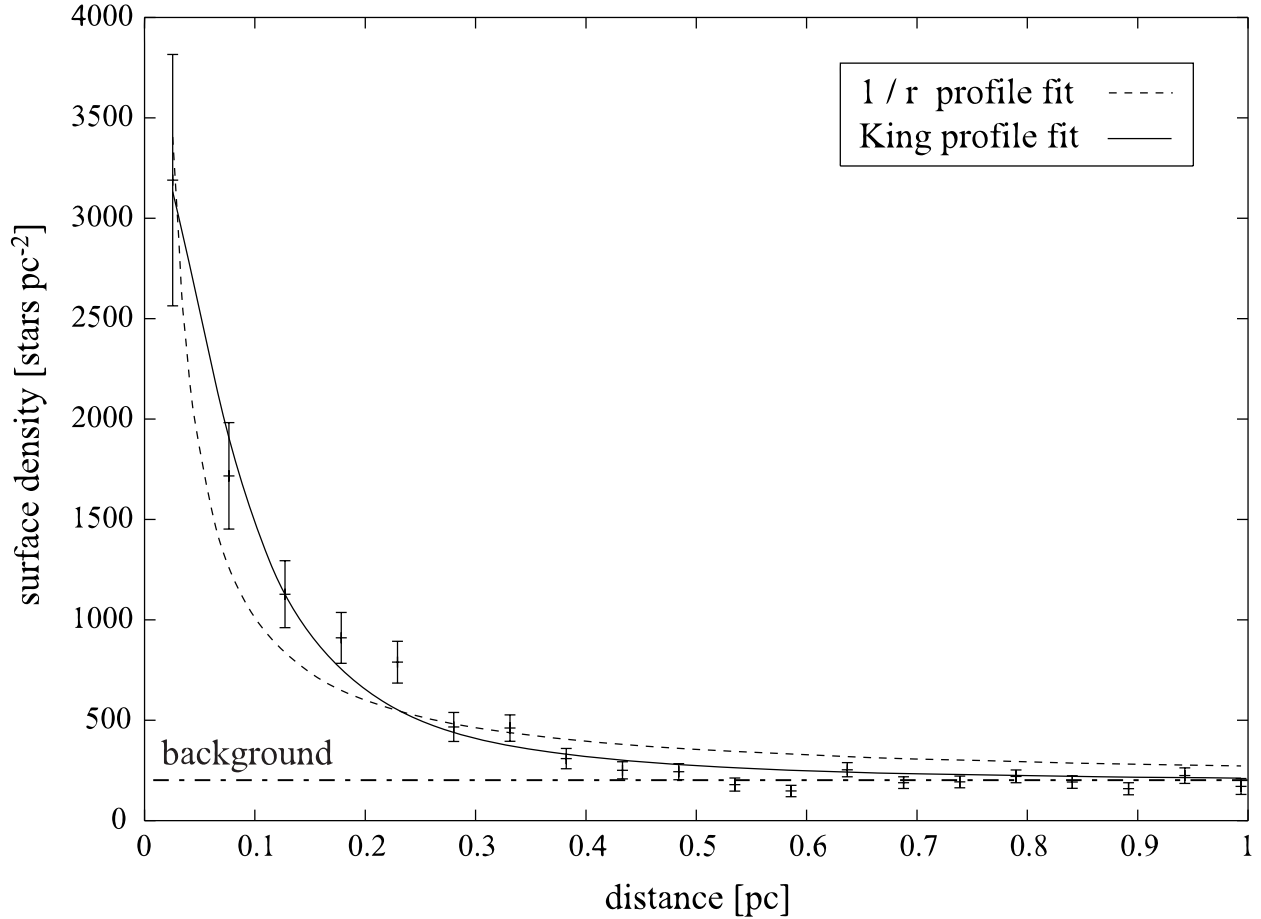}
\caption{Radial profile of the stellar surface number density for the RCW 36 cluster. Stellar surface density (in stars per pc$^{2}$) is measured in circular annuli, centering on the location of the peak position of the stellar surface density (see Fig.~\ref{contour}). Profiles are calculated using annuli with equal radial steps (=15\arcsec). The errors are statistical $\sqrt{N}$ error. The horizonal dot-dashed line corresponds to background surface density [level = 2.0 stars in a 30\arcsec~$\times$ 30\arcsec~unit cell (see \S 3.2.3)]. The solid curve represents a King profile fit, and the dashed curve represents a 1/r profile fit. 
\label{radprof}}
\end{figure}

\clearpage
\begin{figure}
\plotone{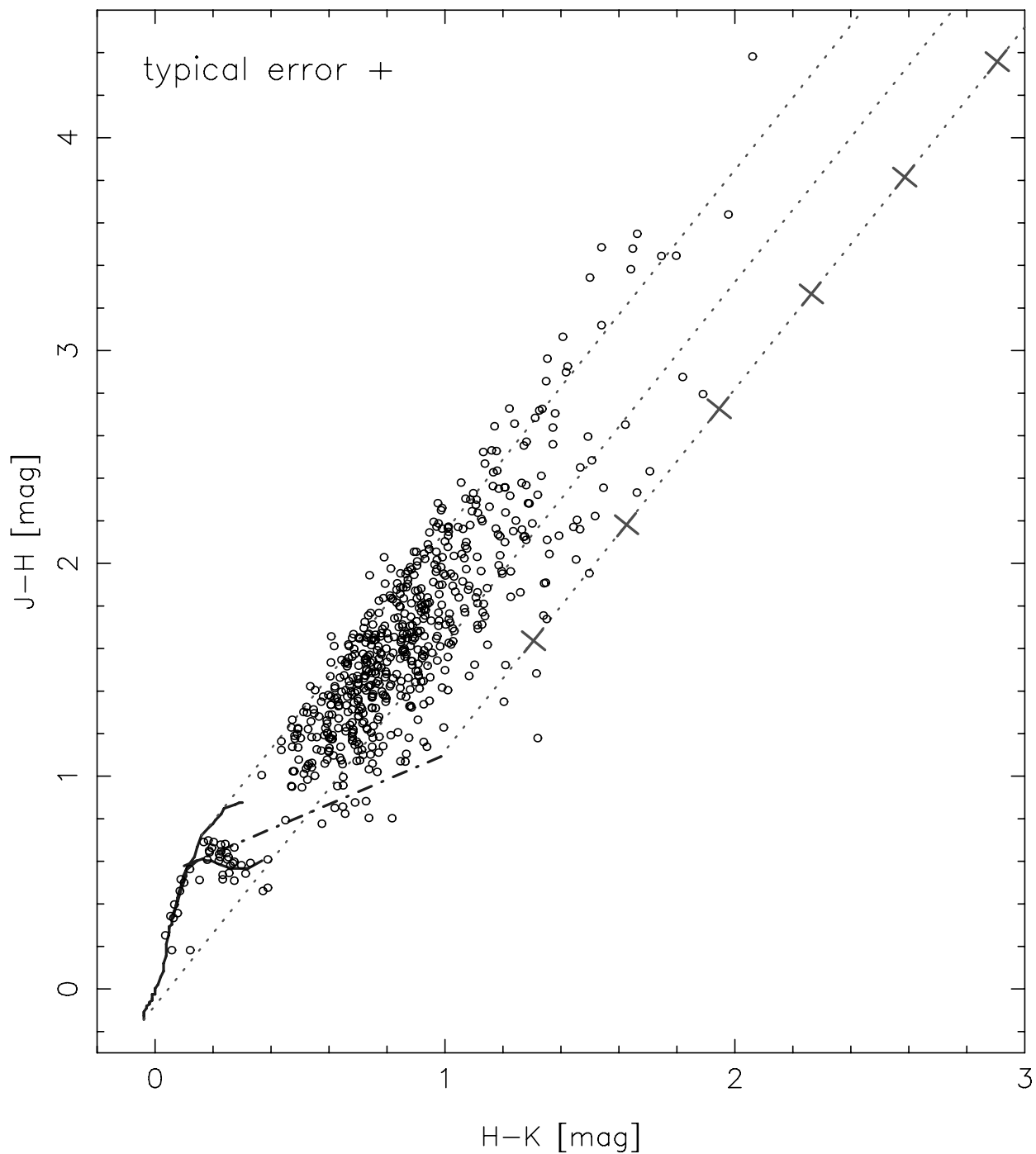}
\caption{The color-color diagram for the sources in the RCW 36 cluster with $J$, $H$, $K_{S}$ band photometric errors $\le$ 0.1 mag. 
The solid lines represent the loci of dwarfs and giants \citep{Tokunaga96}, the dot-dashed line represents the locus of unreddened Classical T Tauri Stars (CTTS) \citep{Meyer97}.
The two leftmost parallel dashed lines define the reddening band for dwarfs and giants. They are parallel to the reddening vector.
The tick marks on the reddening line indicate 5 mag intervals in Av from the right edge of the CTTS locus. The reddening low of \citet{Koornneef83}, having the slope of 1.7, is adopted.
\label{2cd}}
\end{figure}

\clearpage

\begin{figure}
\plottwo{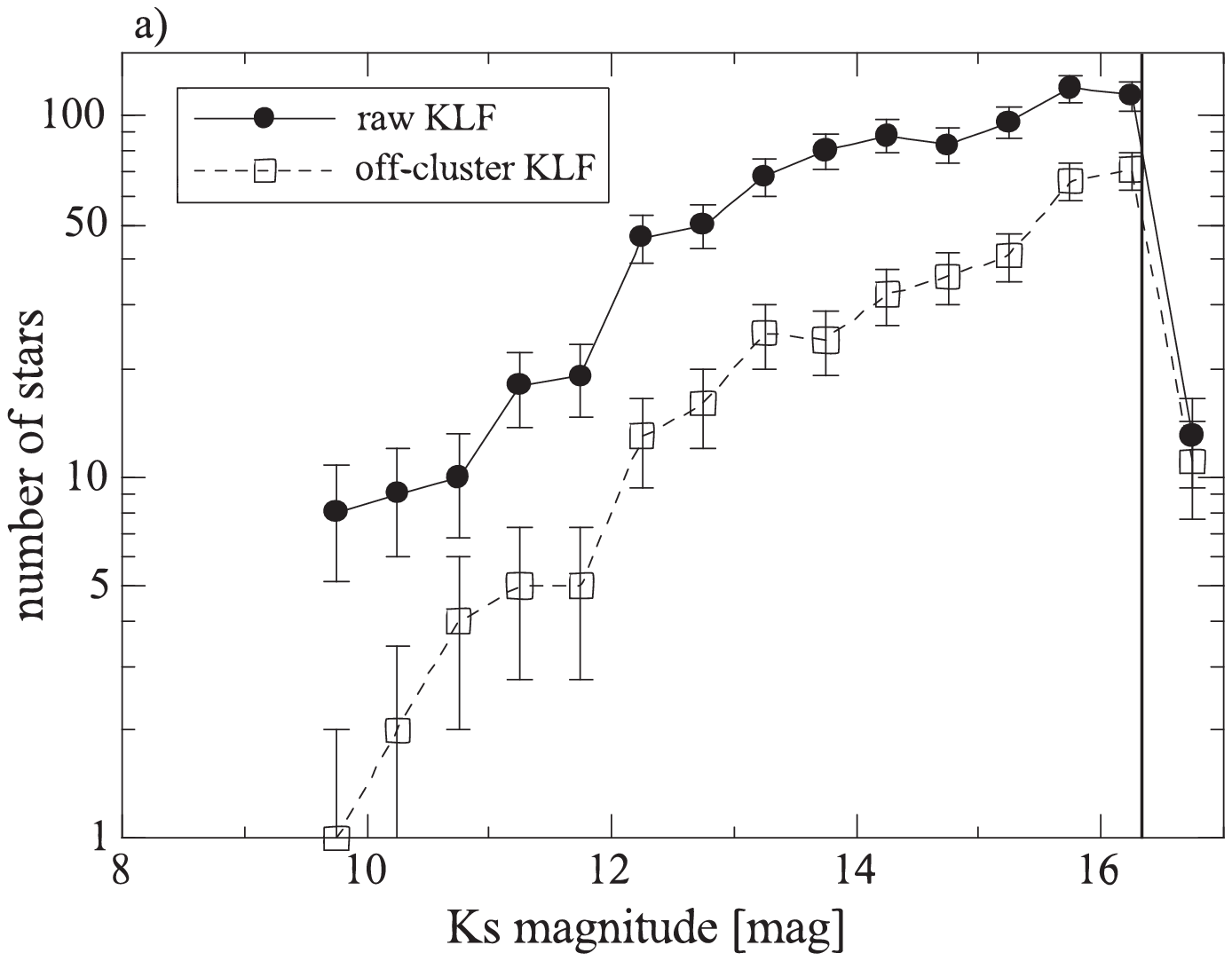}{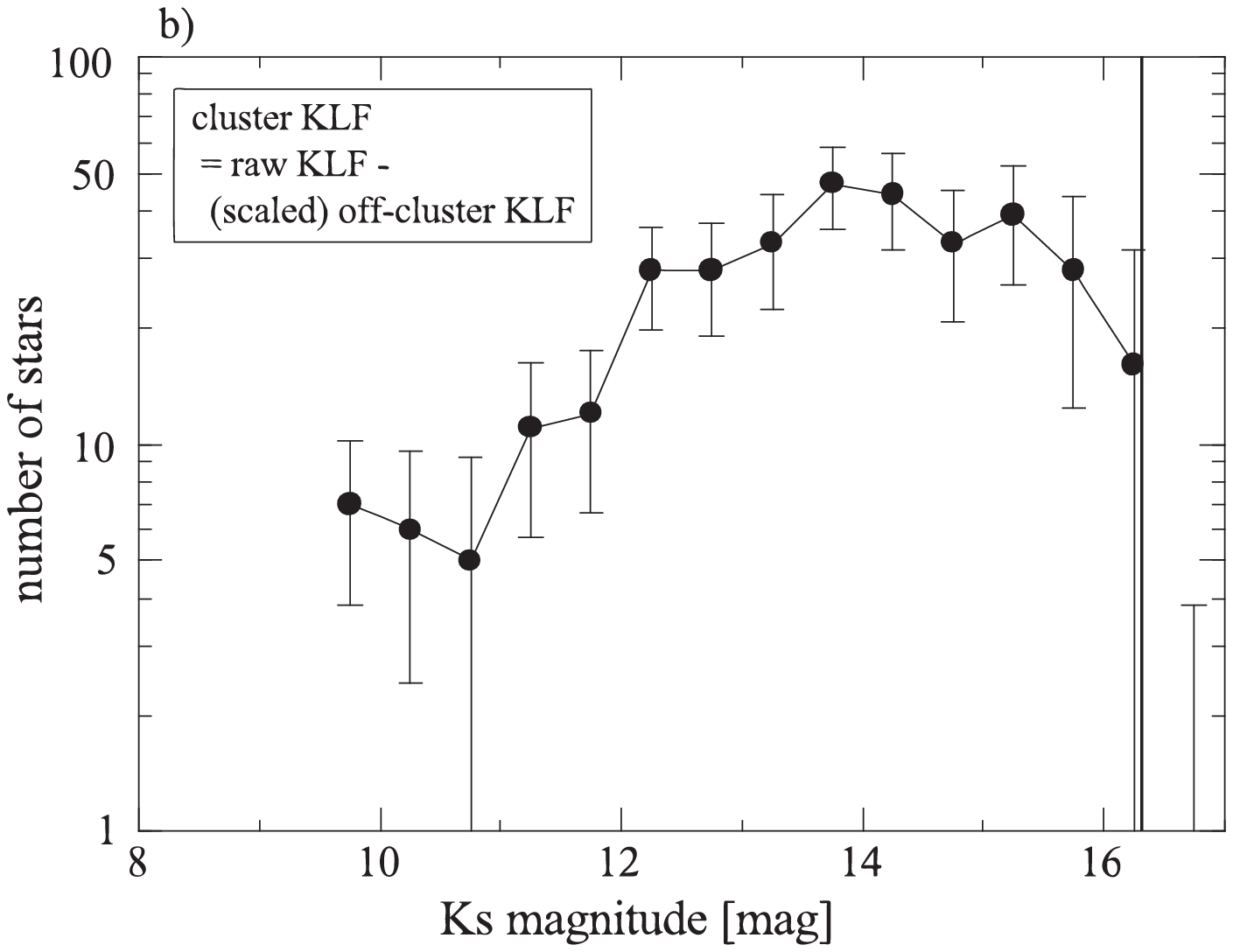}
\caption{The KLFs of the RCW 36 cluster. The bin width is 0.5 mag, and the errors are statistical $\sqrt{N}$ error.
The solid vertical line represents 90 \% completeness limit.
a): The filled circles with solid line are ``raw KLF'' (from all sources detected in the $K_{S}$ band with a photometric error of $\le$ 0.1 mag), and the open squares with dashed line are the ``off-cluster KLF'' [from the sources detected in the $K_{S}$ band  with a photometric error of $\le$ 0.1 mag and are located in the off-cluster region (the region whose SSNDs are $\le$ background + 1 $\sigma$)].
b): The KLF for the RCW 36 cluster members. In order to obtain this KLF, the off-cluster KLF (scaled to account for the different areas) is subtracted from the raw KLF.
\label{KLF}}
\end{figure}

\clearpage
\begin{figure}
\plotone{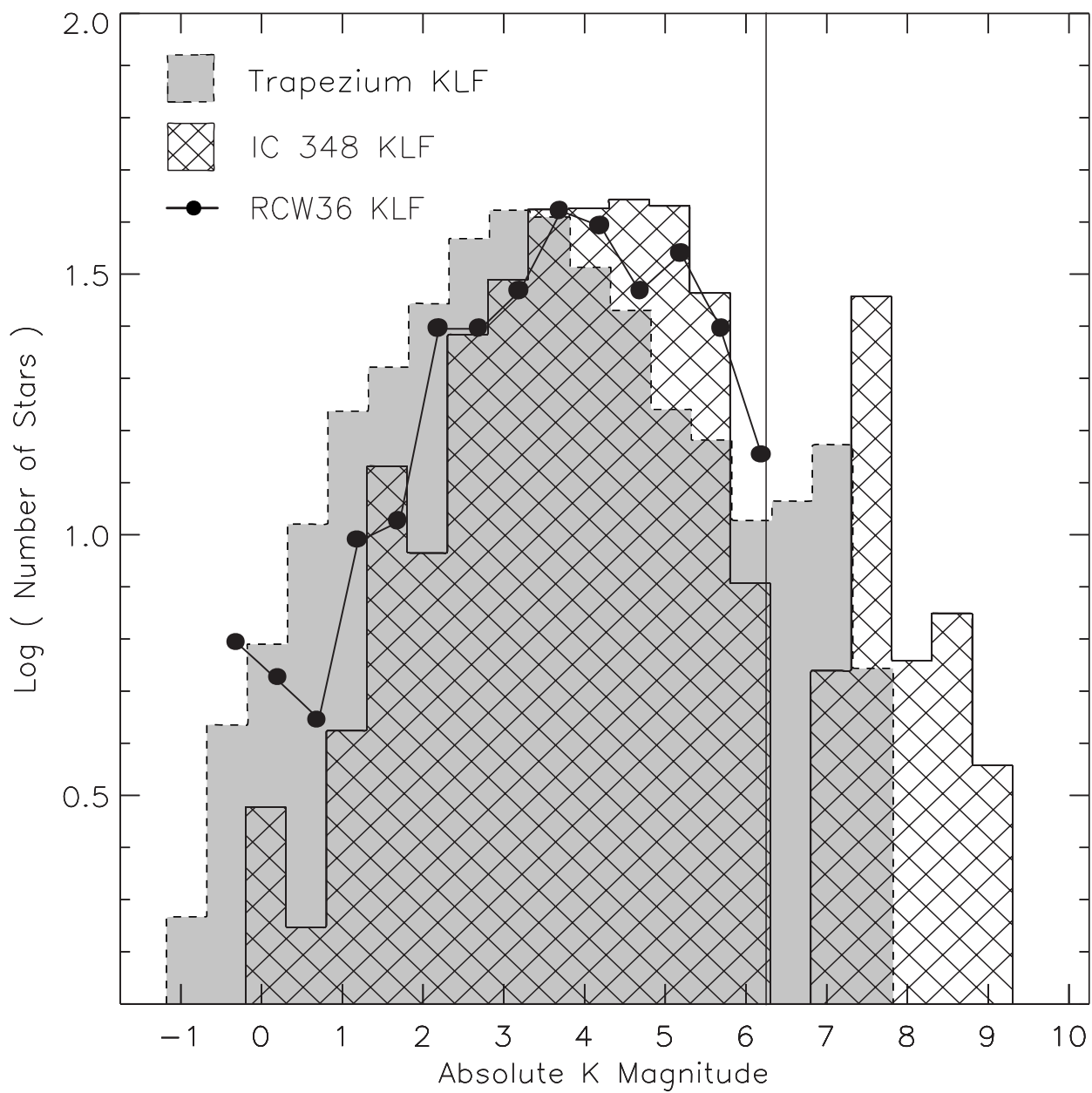}
\caption{The comparisons of the background subtracted KLFs. All magnitudes are shifted to absolute magnitudes. The Trapezium and RCW 36 KLFs have been scaled to contain the same number of stars as IC 348. The KLFs for IC348 and Trapezium are obtained by \citet{Muench03}. The solid vertical line represents our 90 \% completeness limit. Reddening correction was performed by shifting each KLF toward the brighter magnitude by each mean extinction (see \S 4.1).
\label{KLFcomp}}
\end{figure}

\clearpage
\begin{figure}
\plotone{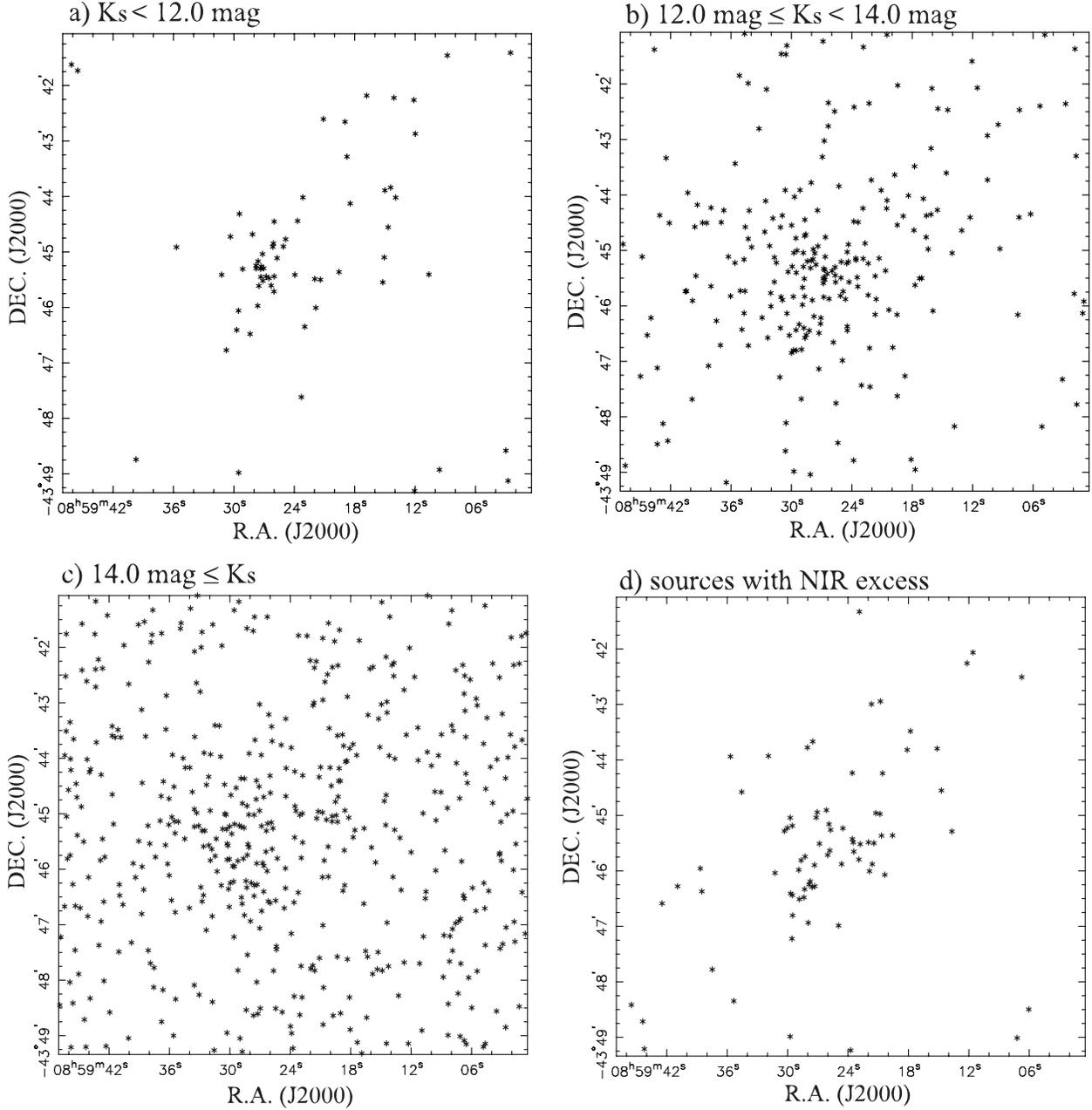}
\caption{a) - c): The distributions of the sources detected in the K$_{S}$ band with a photometric error of $\le$ 0.1 mag [a): the sources of K$_{S} <$ 12.0, b): the sources of 12.0 $\le$ K$_{S} <$ 14.0, c): the sources of 14.0 $\le$ K$_{S}$]. d): The distributions of the sources identified in three bands and falling into a region of infrared excess in Fig.~\ref{2cd}.
\label{masseg_dist}}
\end{figure}

\clearpage
\begin{figure}
\plotone{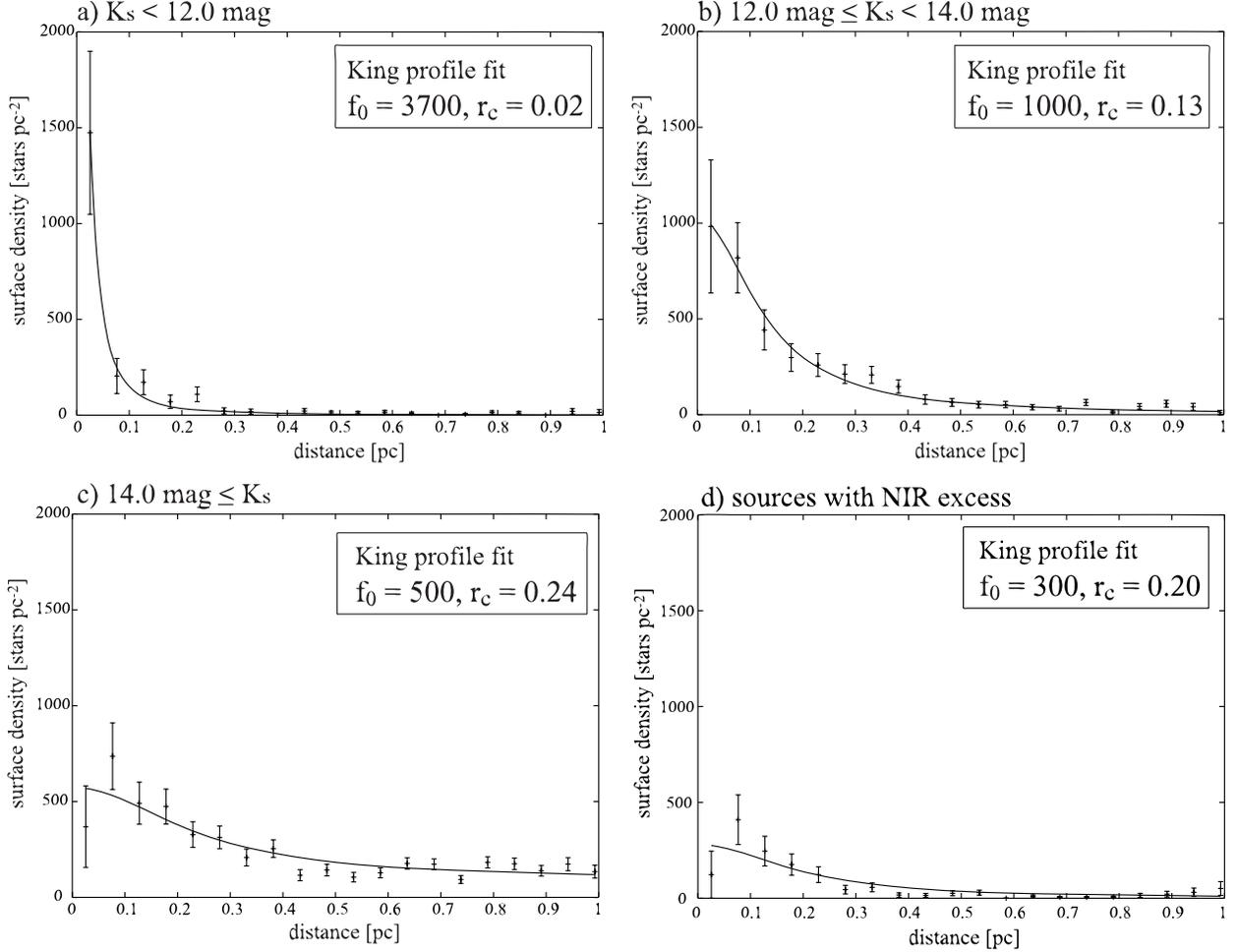}
\caption{Radial profiles of the stellar surface number density for the RCW 36 cluster. The errors are statistical $\sqrt{N}$ error.
%%The horizonaldot-dashed lines represent the backgroud level estimated in \S 3.2.3..
a): the sources of K$_{S} <$ 12.0, b): the sources of 12.0 $\le$ K$_{S} <$ 14.0, c): the sources of 14.0 $\le$ K$_{S}$, d) the sources with NIR excess (see \S 3.3). The photometric errors for all sources are $\le$ 0.1 mag. The solid lines represent King profile fits.
\label{masseg_rad}}
\end{figure}

\clearpage

\begin{deluxetable}{llllll}
\tablecolumns{6}
\tablecaption{Comparison of the most massive star and the central SSND
\label{ssnd_comp}} 
\tablehead{
\colhead{Region} & \colhead{Distance} & \colhead{Most massive} &
\colhead{Central SSND} & \colhead{K} & \colhead{References} \\
\colhead{} & \colhead{[pc]} & \colhead{[Sp type]} &
\colhead{[stars pc$^{-2}$]} & \colhead{(limit)} & \colhead{}
}

\startdata
	R~CrA	& 150 & B9 & 260\tablenotemark{a} & 16.5 & 1 \\
	IC348	& 320 & B5 & 955\tablenotemark{a} & 15.0 & 2 \\
	NGC2024	& 415 & O8 & 1600\tablenotemark{a} & 14.0 & 3, 4, 5 \\
	MonR2	& 830 & B0 & 1900\tablenotemark{c} & 14.5 & 6 \\
	Trapezium & 470 & O7 & 5600\tablenotemark{b}  & 17.0 & 7 \\
	RCW 36	& 700 & O8-O9 & 3000\tablenotemark{a} & 16.3 & this work\\
\enddata

\tablenotetext{a}{innner central 0.1 pc}
\tablenotetext{b}{yielded by King profile fit}
\tablenotetext{c}{yielded by Gaussian profile fit}
\tablerefs{
(1) \citet{Wilking97}; (2) \citet{Lada95}; (3) \citet{Lada91}; (4) \citet{Bik03}; (5) \citet{Haisch00}; (6)\citet{Carpenter97}; (7)\citet{Hillenbrand98}
}
\end{deluxetable}

\clearpage

\begin{deluxetable}{lccccccl}
\tablecolumns{8}
\tablecaption{Estimation of the Relaxation time
\label{relax_comp}} 
\tablehead{
\colhead{Region} & \colhead{Size} & \colhead{$N_{star}$} &
\colhead{$v_{disp}$} & \colhead{$\tau_{cross}$} & \colhead{$\tau_{relax}$} & 
\colhead{Cluster Age} & \colhead{References} \\
\colhead{} & \colhead{[pc]} & \colhead{} &
\colhead{[km s$^{-1}$]} & \colhead{[Myr]} & \colhead{[Myr]} & \colhead{[Myr]} & \colhead{}
}

\startdata
	IC348 & 1.0 & 300 & 1.0\tablenotemark{a} & 2.0 & 10 & 2-3 & 1, 2, 3 \\
	NGC2024 & 0.88 & 309 & 1.3\tablenotemark{b} & 1.3 & 7.1 & 1 & 1, 4, 5 \\
	MonR2 & 1.85 & 371 & 1.1\tablenotemark{b} & 3.3 & 21 & $<$3 & 1, 4, 6 \\
	Trapezium & 0.24 & 780 & 2.5\tablenotemark{c}  & 0.13 & 2.2 & $<$1 & 1, 7, 8 \\
	RCW 36 & 0.5 & 349 & 2.2\tablenotemark{a} & 0.44 & 2.6 & 2-3 & this work, 9 \\
\enddata

\tablenotetext{a}{$\Delta v$ of C$^{18}$O}
\tablenotetext{b}{$\Delta v$ of $^{13}$CO}
\tablenotetext{c}{velocity dispersion of the stars}
\tablerefs{
(1) \citet{Lada03} and references therein; (2) \citet{Bachiller87}; (3) \citet{Haisch01a}; (4) \citet{Miesch94}; (5) \citet{Haisch00}; (6)\citet{Carpenter97}; (7) \citet{Jones88}; (8)\citet{Hillenbrand98}; (9) \citet{Yamaguchi99}
}
\end{deluxetable}

\end{document}